\documentclass[showpacs,preprint]{revtex4}
\usepackage{bm}
\begin{document}
\setcounter{page}{1}
\title
{Determining the size of the proton}
\author
{N. G. Kelkar, F. Garcia Daza and M. Nowakowski}
\affiliation{
Departamento de F\'isica, Universidad de los Andes,
Cra.1E No.18A-10, Santafe de Bogot\'a, Colombia}
\begin{abstract}
A measurement of the Lamb shift of 49,881.88(76) GHz in muonic 
hydrogen in conjunction with theoretical estimates of the 
proton structure effects was recently used to deduce an accurate 
but rather small radius of the proton. Such an important shift in 
the understanding of fundamental values needs reconfirmation. 
Using a different approach with electromagnetic 
form factors of the proton, we obtain a new expression for the transition 
energy, $\Delta = E_{2P_{{3}/{2}}}^{f=2} - E_{2S_{{1}/{2}}}^{f=1}$, 
in muonic hydrogen and deduce a proton radius, $r_p = 0.831$ fm. 
\end{abstract}
\maketitle
\section{Introduction}
In a Quantum Field Theoretic (QFT) description of a muon-proton amplitude, 
the coupling of the photon to a point-like particle (vertex),
$e\gamma_{\mu}$, has to be replaced by $e(F_1(q^2)\gamma_{\mu}+
F_2(q^2)\sigma_{\mu \nu}q^{\nu}/2m_p)$ where $q_{\mu}$ is the four momentum
carried by the photon, 
$q^2=q_{\mu}q^{\mu}$ and $F_i(q^2)$ are the electromagnetic
form-factors \cite{bosted,others}
which encode the information on the structure of the proton. 
The inclusion of form factors introduces corrections to the binding energy 
in Coulomb bound systems such as the hydrogen atom. In fact a systematic 
inclusion of the proton structure effects introduces corrections to other 
terms such as the 
fine and hyperfine splittings in the hydrogen atom \cite{we3}. 
Given the recent high
precision of experimental measurements, the corrections due to the proton size 
become relevant. One such precision measurement using pulsed laser 
spectroscopy was recently performed \cite{pohl} to measure the 
Lamb shift in muonic hydrogen ($\mu^- p$). Using this accurate measurement of 
the energy difference (Lamb shift) between the $2S_{1/2}$ and $2P_{3/2}$ 
states, a precise value for the transition energy, 
$\Delta = E_{2P_{{3}/{2}}}^{f=2}
-E_{2S_{{1}/{2}}}^{f=1} = 206.2949(32)$ meV, in muonic hydrogen was given. 
Comparing the measured value with a model calculation of this difference 
which included the proton structure effects, a new value of the proton 
radius, namely, $r_p = 0.84184(67)$ fm was published. It was further claimed 
that the new value implies that either the Rydberg constant has to be shifted 
by -$110$ kHz/c or the calculations of the QED effects in atomic hydrogen 
or muonic atom are insufficient. 
The uncertainties due to form factors have been discussed 
in \cite{pohl}. 
Another uncertainty however seems 
to arise from the way in which one incorporates the finite size corrections 
(FSC) due to the structure of the proton. The FSC 
in \cite{pohl} were based on methods given in \cite{zemach,friar}. 
Here we implement the FSC based on the Breit equation \cite{we3,wee1} and 
thus obtain a different form 
for the expression of the energy difference, 
namely, $\Delta (= E_{2P_{3/2}}^{f=2} - E_{2S_{1/2}}^{f=1})  
= 209.16073 + 0.1139 r_p - 4.3029 r_p^2 + 0.02059 r_p^3 \, \, {\rm meV}$, 
as compared to, 
$\Delta^{Nature} 
= 209.9779(49) - 5.2262 r_p^2 + 0.0347 r_p^3$ meV, 
obtained in \cite{pohl}. Based on this new expression and following the same 
procedure as in \cite{pohl}, we extract a new radius of the proton. 
This value of $r_p = 0.83112$ fm based on the FSC from the Breit 
equation is not far from that predicted in \cite{pohl}, namely 
$r_p=0.84184(67)$. However, given 
the precision of the numbers under consideration, it is worth noting that 
the method used for proton structure corrections does introduce 
an uncertainty.

Before proceeding to the formalism used in the present work, we note 
certain points which motivated the present work. 
The proton radius is 
one of the fundamental numbers in nuclear and particle physics and is 
obviously relevant for atomic physics too. The radius estimated in \cite{pohl} 
is smaller than the value obtained by some other methods (see 
\cite{pohl} for references and discussions on this issue). This could mean 
that some corrections to the energies in the hydrogen atom are missing. 
If a shift in the understanding of the proton radius occurs, it is necessary 
to reconfirm the result with an independent calculation. This is 
particularly so in view of the model dependence entering the extraction of 
the radius. The nuclear physics inputs (here, the structure of the proton) 
are never as precise as their electroweak counterparts. In the present work, 
we try to provide this reconfirmation using a different theoretical 
approach for the proton structure corrections. Though we do agree 
qualitatively on a smaller radius, we notice that the accuracy gets  
blurred by the approach used for finite size corrections.

\section{Breit potential with form factors}
All $\vec{r}$
dependent potentials in Quantum Field Theory (QFT) are obtained
by Fourier transforming an elastic scattering amplitude
suitably expanded in $1/c^2$ (see \cite{we3} for several examples). 
The Breit equation \cite{bethe,LLbook} 
follows the very same principle for elastic
$e^-\mu^+$, $e^+e^-$ (positronium), $e^-p$ (hydrogen) and
$\mu^-p$ (muonic hydrogen) amplitudes.
The one-photon exchange amplitude between
the proton and the muon leads then to the Coulomb potential plus the fine
and hyperfine structure (hfs), the Darwin term and the retarded potentials 
\cite{bethe,LLbook}. Here we use a modified Breit potential \cite{we3} for the 
$\mu^- p$ system which includes the 
electromagnetic form factors of the proton. 
We present a calculation based on 
this potential with form factors to evaluate the transition energy,  
$\Delta$, and hence the 
proton radius, $r_p$, as in \cite{pohl}.
The Fourier transform of the momentum space Breit potential in \cite{we3} 
gives $V = -\alpha/r + \delta V(\hat{p}_{\mu}, \hat{p}_p, r)$, where 
$\delta V(\hat{p}_{\mu}, \hat{p}_p, r)$ contains Coulomb, Darwin, fine and 
hyperfine terms with form factors. 
The expectation values of these terms give the energy corrections to the 
various terms. 

The standard Breit potential is normally written down either for
point-like particles (with standard point-like vertices) or
at zero momentum transfer at the vertices (i.e. $F_1(0)$ and $F_2(0)$), at 
least for the hadronic vertex. Therefore, extending it to include the
full $q^2$ dependence of the proton form factors (which encodes the
finite size corrections) is just a straightforward
and mild extension of the standard procedure. Thus one can obtain the 
Breit potential in momentum space for the $e^- p$ or $\mu^- p$ system 
with the proton structure effects fully included and is given as 
(Eq. (21) in \cite{we3}): 
\begin{eqnarray}\label{potinqspace} 
&&\hat{U}(\textbf{p}_X,\textbf{p}_p,\textbf{q})=4\pi e^2\Bigg[F_1^XF_1^p\Bigg
(-\frac{1}{\textbf{q}^2}+\frac{1}{8m_X^2c^2}
+\frac{1}{8m_p^2c^2}+\frac{i\bm{\sigma}_p.(\textbf{q}\times\textbf{p}_p)}
{4m_p^2c^2\textbf{q}^2}
-\frac{i\bm{\sigma}_X.(\textbf{q}\times\textbf{p}_X)}{4m_X^2c^2\textbf{q}^2}	\nonumber\\\nonumber\\	& &
 +\frac{\textbf{p}_X.\textbf{p}_p}{m_X m_p c^2\textbf{q}^2}-\frac
{(\textbf{p}_X.\textbf{q})(\textbf{p}_p.\textbf{q})}{m_X m_p c^2\textbf{q}^4}
-\frac{i\bm{\sigma}_p.(\textbf{q}\times\textbf{p}_X)}{2m_X m_pc^2\textbf{q}^2}
+\frac{i\bm{\sigma}_X.(\textbf{q}\times\textbf{p}_p)}
{2m_X m_pc^2\textbf{q}^2}+\frac{\bm{\sigma}_X.\bm{\sigma}_p}{4m_X m_pc^2}	
\nonumber\\\nonumber\\& &
 -\frac{(\bm{\sigma}_X.\textbf{q})(\bm{\sigma}_p.\textbf{q})}
{4m_X m_pc^2\textbf{q}^2}\Bigg)+F_1^X F_2^p\Bigg(\frac{1}{4m_p^2c^2}
+ \frac{i\bm{\sigma}_p.(\textbf{q}\times\textbf{p}_p)}{2m_p^2c^2\textbf{q}^2}  
- \frac{i\bm{\sigma}_p.(\textbf{q}\times\textbf{p}_X)}{2m_X m_pc^2\textbf{q}^2}
-\frac{(\bm{\sigma}_X.\textbf{q})(\bm{\sigma}_p.\textbf{q})}{4m_X m_pc^2
\textbf{q}^2} 	
\nonumber\\\nonumber\\& &+ \frac{\bm{\sigma}_X.\bm{\sigma}_p}{4m_X m_pc^2}\Bigg)
 +F_2^X F_1^p\Bigg(\frac{1}{4m_X^2c^2} - \frac{i\bm{\sigma}_X.
(\textbf{q}\times\textbf{p}_X)}{2m_X^2c^2\textbf{q}^2}  
+ \frac{i\bm{\sigma}_X.(\textbf{q}\times\textbf{p}_p)}{2m_X m_pc^2\textbf{q}^2}
-\frac{(\bm{\sigma}_X.
\textbf{q})(\bm{\sigma}_p.\textbf{q})}{4m_X m_pc^2\textbf{q}^2} \nonumber\\\nonumber\\& &	
+ \frac{\bm{\sigma}_X.\bm{\sigma}_p}{4m_X m_pc^2}\Bigg)+	
F_2^X F_2^p\Bigg(\frac{\bm{\sigma}_X.\bm{\sigma}_p}{4m_X m_pc^2}
-\frac{(\bm{\sigma}_X.\textbf{q})(\bm{\sigma}_p.\textbf{q})}{4m_X m_pc^2
\textbf{q}^2}\Bigg)
 \Bigg] ,\label{potFF}
\end{eqnarray}
where, $X = e$ or $\mu$. 
The Fourier transform of this potential results in a
space dependent potential suitable for calculations of the corrections
to the atomic energy levels via time-independent perturbation theory.

The two form factors, $F_1(q^2)$ and $F_2(q^2)$ of the proton are connected 
to the Sachs form factors, $G_E(q^2)$ and $G_M(q^2)$. The latter can be 
interpreted in the proton rest frame to be Fourier transforms of the 
charge and magnetization distributions in the proton. 
They can be approximated fairly well by a dipole form 
\cite{bosted} (which is suitable for analytic calculations) 
as, $G_D({\bf q}^2) = 1/(1 + {\bf q}^2/m^2)^2 \approx G_E^p({\bf q}^2) \approx 
G_M^p({\bf q}^2) / \mu_p$,  
where ($1\, + \,\kappa_p$)
= $\mu_p$ = 2.793 is the proton's 
magnetic moment and $m$ the dipole parameter. We use the 
standard non-relativistic approximation $q^2 \approx - {\bf q}^2$ to 
obtain the dipole form of $F_1^p({\bf q}^2)$ and $F_2^p({\bf q}^2)$ 
(see Eq.(30) in \cite{we3}). 
The parameter $m$ is eventually related to 
the proton radius and taken as a free parameter to fix the radius of the 
proton (to be discussed below). 

\section{Proton structure corrections to energies}
We follow a procedure similar to that in \cite{pohl} to obtain the 
proton radius with the difference that the proton structure corrections 
are included via the Breit potential method. The Breit potential $V(r)$ as 
explained above consists of a series of terms \cite{bethe} 
with the leading term being 
the one due to the Coulomb potential. Thus, the expectation values of 
$V(r) = V_C + V_D + V_{hfs} + V_{SO} + ... $ give rise to the various 
energy terms including the standard Coulomb, Darwin, fine structure, 
hyperfine structure etc. Below, we shall give the expression for some of 
the energies (using the 
Fourier transform of the momentum space potential with form factors \cite{we3}) 
which are relevant for the calculation of the radius. For details of the 
full potential in coordinate space, we refer the reader to \cite{we3}. 
\subsection{Coulomb term} 
The first term, namely, the Coulomb potential, $\hat{V}_C^{FF}$, 
with form factors is given as 
$\hat{V}_C^{FF} ={V}_C+\Delta {V}_C$, i.e., 
\begin{equation}\label{coulpot}
\hat{V}_C^{FF}=-\frac{e^2}{r}
\Bigg[1-\Bigg(1+\frac{\kappa_p}{(1-k^2)^2}\Bigg)e^{-mr}-\Bigg(1+\frac{\kappa_p}
{(1-k^2)}\Bigg)\frac{m}{2}re^{-mr}+\frac{\kappa_p}{(1-k^2)^2}
e^{-mkr}\Bigg],
\end{equation}
where $k = 2 m_p /m$ with $m_p$ 
being the mass of the proton and $m$ the parameter 
entering the dipole form factor. Note that the above expression has been 
obtained under the condition that $2 m_p \ne m$ and hence does not generate 
any singularities (which as such should not be expected too since the 
form factors are not singular). 
A small explanation regarding 
this and the other analytic expressions which follow is in order here.
The potential in coordinate space is obtained by Fourier transforming the 
potential in momentum space. If we consider the Coulomb term (one with 
${F_1^X\, F_1^p(Q^2)/ Q^2}$ in Eq.(\ref{potinqspace}), where we write 
$Q^2 = {\bf q}^2$), Fourier transform it and replace $F_1^p(Q^2)$ using the 
dipole form factor, we obtain
\begin{equation}\label{potr}
\hat{V}_C^{FF}(r) = -{2 e^2\over \pi r} \, \biggl [ \int_0^{\infty} dQ\,
{\sin(Qr) m^4 \over Q (m^2 + Q^2)^2} \, +\, \kappa_p \int_0^{\infty} dQ\, 
{Q \,\sin(Qr) \,m^4 \over (m^2+Q^2)^2(4m_p^2+ Q^2)}\, \biggr]\, .
\end{equation}
We now use a partial fraction expansion to write the terms in the integrands.
As an example, consider the second term in the square bracket of 
Eq.(\ref{potr}). We can see that
$${1 \over (m^2 + Q^2)^2 (4 m_p^2 + Q^2)} = {1 \over (4m_p^2 - m^2)^2 
(4m_p^2 + Q^2)} \,- \,{1\over (4m_p^2 - m^2)^2 (m^2 + Q^2)} $$
$$\quad + {1 \over 
(4m_p^2 - m^2) (m^2 + Q^2)^2}$$
assuming of course that $4 m_p ^2 \ne m^2$. The integrals in (\ref{potr}) can
be performed analytically and lead to Eq.(\ref{coulpot})  
which contains terms of the type
$${1\over4m_p^2 - m^2} = - {1\over m^2} {1 \over  (1 - k^2)}$$
with $k = 2m_p/m$. 
Should it happen that $4m_p^2 = m^2$, then
$1/ [(m^2 + Q^2)^2 (4 m_p^2 + Q^2)] = [1 /(m^2 + Q^2)^3]$
and the integral $\int_0^{\infty} dQ \,[Q \, \sin(Qr)/ 
(m^2 + Q^2)^3]$ can be evaluated by differentiating twice the integral 
$I_1 = \int_0^{\infty} dQ\,[ Q \, \sin(Qr) / (m^2 + Q^2)] = (\pi/ 2)
\, e^{-mr}$ with respect to $m$. However, the case of $4m_p^2 = m^2$ seems 
unrealistic and is not considered in the present work. 

The correction to the Coulomb energy due to form factors is then found as, 
$\Delta E_C=\langle nlm|\Delta{V}_C|nlm\rangle$, which with $\Delta{V}_C$ 
being only a function of $r$ becomes,  
$\Delta E_C = \int_0^{\infty}\Delta {V}_C |R_{nl}(r)|^2r^2dr$ 
(with $R_{nl}(r)$ being the unperturbed hydrogen atom radial function as found 
in books). The correction to the Coulomb energy for any $n, l$ is 
thus given by a lengthy expression involving hypergeometric functions 
(as explained in the appendix). Performing a series expansion of the 
hypergeometric function and truncating it at large orders of the fine structure 
constant $\alpha$, we evaluate the correction to the energy in terms of the 
dipole parameter $m$. 
In order to rewrite the corrections in terms of the proton radius $r_p$, 
we perform an expansion of the type $G_E^p(q^2) = 1 - 2q^2/m^2 + ...$ 
of the Sachs dipole form factor and 
compare it with the standard $G_E^p(q^2) = 1 - <r^2> q^2/6 + ...$. One 
can then write $<r^2> = 12/m^2$ and convert the corrections 
for energies given in terms of $m^2$ 
to those in terms of $<r^2>$. For convenience, in what follows, we shall 
denote $<r^2>^{1/2} = r_p$, $<r^2> = r_p^2$, $<r^2>^{3/2} = r_p^3$ etc. 
Thus the corrections to the Coulomb terms 
of the $2S_{1/2}$ and $2P_{1/2}$ levels 
used in the present work can be rewritten in terms of the 
proton radius $r_p$ as follows:
\begin{eqnarray}\label{coulombnew0}
&&\Delta E_{Coul}^{2S_{1/2}} = {m_r^3 \alpha^4 \over 24} \, 
(A_1 - A_2 + A_3)  \, r_p^2 \, + 
{m_r^4 \alpha^5 \over 12 \sqrt{12} } (-2 A_1 + 2 {A_2 \over k} - 3 A_3) r_p^3  
\\ \nonumber
&&+ {21 m_r^5 \alpha^6 \over 4} ( A_1 - { A_2 \over k^2} + 2 A_3) {r_p^4 
\over 144} \,+ \ldots , 
\end{eqnarray}
with, 
$A_1 = 1 + [{\kappa_p / (1 - k^2)^2}]$, $A_2 = {\kappa_p/ [k^2 
(1 - k^2)^2]}$, 
$A_3 = 1  + [{\kappa_p / (1 - k^2)}]$ and 
$k = {2 m_p/ m}$. 
\begin{equation}
\Delta E_{Coul}^{2P_{1/2}}={m_r^5 \alpha^6 \over 4}\, A\, {r_p^4 \over 144} 
\,- \, {m_r^6 \alpha^7 \over 2} \, B\, {r_p^5 \over (12)^{5/2}}\, + \ldots, 
\end{equation}
where, $A = 3 + [\kappa_p/(1 - k^2)^2] + [2\kappa_p/(1 - k^2)] 
- [1/(k^4(1-k^2)^2)] $ and $B = 7 + [2\kappa_p/(1 - k^2)^2] + 
[5 \kappa_p/(1-k^2)] - [2/(k^5(1-k^2)^2)$, makes a negligible contribution to
the finite size corrections. 
The Coulomb terms of the present work and those 
used in \cite{pohl} will be compared in detail in section IV.
\subsection{Darwin term} 
Let us first look at the Darwin term in the standard Breit equation. The 
potential in coordinate space is given as:
\begin{equation}
  \hat{V}_D^{{\rm no} \,F_{1,2}(q^2)}
=\frac{\pi e²}{2m_X^2c²}\delta(\textbf{r})\Bigg[1+\frac{m_X^2}{m_p^2}\Bigg].
\end{equation}
With the inclusion of the electromagnetic form factors, 
\begin{equation}\label{darwinpot} 
 \hat{V}_D^{F_{1,2}(q^2)}=\frac{e^2}{8m_X^2c^2}\Bigg[(1+2\kappa_X) {G}_{1} +\frac{m_X^2}{m_p^2} {G}_{2}\Bigg], 
\end{equation}
which in the static limit, i.e. taking $F_{1,2}^p(q^2=0)$ reduces to
\begin{equation}
 \hat{V}_D^{F_{1,2}(q^2=0)}=\frac{\pi e^2}{2m_X^2c^2}\delta(\textbf{r})\Bigg[1+2\kappa_X+\frac{m_X^2}{m_p^2}(1+2\kappa_p)\Bigg].
\end{equation}
Here $X = e$ or $\mu$ depending on if we are considering the usual hydrogen 
atom or muonic hydrogen atom. 
In the above, 
\begin{eqnarray}
{G}_{1} &=&\Bigg(1+\frac{\kappa_p}{1-k²}\Bigg)\frac{m³}{2}e^{-mr}+\frac{m²k²\kappa_p}{(1-k^2)^2}\frac{e^{-mr}}{r}-\frac{m²k²\kappa_p}{(1-k^2)^2}
\frac{e^{-mkr}}{r},\nonumber\\
{G}_{2} &=&\Bigg(1+\kappa_p\left(\frac{1-2k²}{1-k²}\right)\Bigg)\frac{m³}{2}e^{-mr}-\frac{m²k²\kappa_p}{(1-k²)^2}\frac{e^{-mr}}{r}+\frac{
m²k²\kappa_p}{(1-k²)^2}\frac{e^{-mkr}}{r}.\nonumber
\end{eqnarray}
Taking the expectation values of these potentials using first order 
time-independent perturbation theory, the corrections to the energies 
corresponding to the Darwin term can be found. 
For example, the Darwin term without form factors and for $l = 0$ is given as, 
\begin{equation}\label{darwin1} 
E_{D}^{{\rm no}\, F_{1,2}(q^2)}(n,l=0) = \frac{\alpha}{2m_X^2c^2}\frac{1}{n^3a_r^3}\Bigg[1+\frac{m_X^2}{m_p^2}\:\:\Bigg] ,\label{4_5}
\end{equation}
and one with the inclusion of $q^2$ dependent proton form factors is 
\begin{equation}\label{darwin3}
E_D^{F_{1,2}(q^2)}(n,l)=\frac{\alpha}{2m_X^2c^2}\frac{1}{n^3a_r^3} \Bigg[(1+2\kappa_X)G_{D1}(n,l) +\frac{m_X^2}{m_p^2}G_{D2}(n,l)\Bigg],
\end{equation}
where $G_{D1}(n,l)$ and $G_{D2}(n,l)$ are lengthy expressions 
involving the hypergeometric functions, $_2F_1$. The Bohr radius,  
$a_r = 1/ (m_r \alpha)$, where $m_r$ is the reduced mass. 
If we restrict to form factors taken 
at $q^2 = 0$, then, 
\begin{equation}\label{darwin2} 
E_{D}^{F_{1,2}(q^2=0)}(n,l=0)=\frac{\alpha}{2m_X^2c^2}\frac{1}{n^3a_r^3}
\Bigg[1+2\kappa_X+\frac{m_X^2}{m_p^2}\:\:(1+2\kappa_p)\Bigg].\label{4_6}
\end{equation}

In order to get a better insight into the expressions, we 
rather replace the hypergeometric function $_2F_1$ by its series expansion 
$_2F_1 (a,b;c;z) \,=\,1\, +\, (ab/c)(z/1!)\, +\, 
[a (a+1) b (b+1)]/[c(c+1)](z^2/2!)\, +\, ...... $ in the expressions 
for energies and truncate the series at large 
orders of the fine structure constant $\alpha$. Substituting for the dipole 
parameter by $m^2 = 12/r_p^2$, the corrections to the energies corresponding 
to the Darwin terms with $n=2$ and $l=0,1$ are given as
\begin{eqnarray}
E_{D}^{F_{1,2}(q^2)}(2S_{1/2})=\frac{m_{r}^3}{m_X^2}\frac{\alpha^4}{2^4}
\Bigg[\left((1+2\kappa_X)+\frac{m_X^2}{m_p^2}(1 + 2\kappa_p)\right)+
\left((1+2\kappa_X)A_1+\frac{m_X^2}{m_p^2}
B_1\right)r_p\\ \nonumber
+\left((1+2\kappa_X)A_2+\frac{m_X^2}{m_p^2}
B_2\right)r_p^{2} + \left((1+2\kappa_X)A_3+\frac{m_X^2}{m_p^2}
B_3\right)r_p^{3}+\ldots\Bigg]
\end{eqnarray}
and
\begin{equation}
E_{D}^{F_{1,2}(q^2)}(2P_{1/2})=\frac{m_{r}^3}{m_X^2}\frac{\alpha^4}{2^4}
\Bigg[\left((1+2\kappa_X)D_2+\frac{m_X^2}{m_p^2}
E_2\right)r_p^{2}+\left((1+2\kappa_X)D_3+\frac{m_X^2}{m_p^2}
E_3\right)r_p^{3}+\ldots\Bigg].
\end{equation}
The coefficients, $A_1$, $B_1$ etc in the above equations depend on $\kappa_p$ 
and $k = 2 m_p /m$. The correction for the $2P$ state turns out to be 
negligibly small and is not of 
much relevance for the present work. The coefficients which contribute 
to the $2S$ energy correction up to order $r_p^2$ are given as 
\begin{eqnarray*}
A_1&=&\frac{1}{a_r\sqrt{12}}\left[-6\left(1+\frac{\kappa_p}{1-k^2}\right)-4\frac{k^2\kappa_p}{(1-k^2)^2}+4\frac{\kappa_p}{k(1-k^2)^2}\right]\\ 
A_2&=&\frac{1}{48a_r^{2}}\left[84\left(1+\frac{\kappa_p}{1-k^2}\right)+42\frac{k^2\kappa_p}{(1-k^2)^2}-42\frac{\kappa_p}{k^{2}(1-k^2)^2}\right]\\ 
B_1&=&\frac{1}{a_r\sqrt{12}}\left[-6\left(1+\kappa_p\frac{1-2k^2}{1-k^2}\right)+4\frac{k^2\kappa_p}{(1-k^2)^2}-\frac{\kappa_p}{k(1-k^2)^2}  
\right]\\ 
B_2&=&\frac{1}{48a_r^{2}}\left[ 84\left(1+\kappa_p\frac{1-2k^2}{1-k^2}\right)-42\frac{k^2\kappa_p}{(1-k^2)^2}+42\frac{\kappa_p}{k^{2}(1-k^2)^2} 
\right]. 
\end{eqnarray*}
Here, $a_r = 1/(m_r \alpha)$. 

\subsection{Fine structure}
As mentioned in the beginning, the expectation values of the 
various terms in the 
potential in q-space, Eq. (\ref{potinqspace}), give rise to the Coulomb,
Darwin, retarded potential, fine and hyperfine structures in the hydrogen 
atom. The fifth and the ninth spin-orbit terms in this equation are the 
ones corresponding to the fine structure. A Fourier transform of this 
potential leads to the following potential for the fine structure 
with form factors: 
\begin{equation}
\hat{V}_{FS}=\frac{\alpha}{4m_X^2c^2}\Bigg[(1+2\kappa_X)+\frac{2m_X}{m_p}
(1+\kappa_X)\Bigg]\left(\frac{1}{r^3}+\frac{{G}_{FS}}{r^3}\right)\textbf{L}.\textbf{S}_X,
\end{equation}
where, 
\begin{eqnarray} 
{G}_{FS}&=&-\bigg(1+\frac{\kappa_p}{(1-k²)^2}\bigg)e^{-mr}(1+mr)-\bigg(1+\frac{\kappa_p}{(1-k²)}\bigg)	\frac{m²}{2}r^2e^{-mr}\nonumber\\&&+\frac{\kappa_p}{(1-k²)^2}e^{-mkr}(1+mkr).\nonumber
\end{eqnarray}
Using first order time-independent 
perturbation theory for evaluating the expectation value 
of the above potential, the fine structure energy term with the effect of 
form factors is given as, 
\begin{eqnarray}\label{fineenergy} 
E_{FS}(n,l,j)&=&\frac{\alpha}{4m_X^2c^2}\Bigg[(1+2\kappa_X)+
\frac{2m_X}{m_p}(1+\kappa_X)\Bigg]\bigg(j(j+1)-l(l+1)-s_X(s_X+1)\bigg)
\nonumber\\&&
 \Bigg(\frac{1}{l(l+1)(l+\frac{1}{2})n^3a_r^3}+G_{FS}(n,l)\Bigg),
\end{eqnarray}
where, $G_{FS}(n,l)$ is given by a lengthy expression which 
contains the Gamma functions and 
hypergeometric functions. The electron (or muon) 
total angular momentum $j = l + s_X$, where $l$ is the orbital angular 
momentum and $s_X$ the spin (with $X = e$ or $\mu$). 
The fine structure levels relevant for the present work are the 
$2P_{3/2}$ and $2P_{1/2}$ levels which appear in the difference, 
$\Delta E_{FS}^{2P_{3/2}} = E_{FS}^{2P_{3/2}} - E_{FS}^{2P_{1/2}}$. 
The proton structure corrections to the $2P$ levels are always much 
smaller than those to the $2S$ levels, however, for completeness we 
give these expressions below. 
\begin{equation}
 E_{FS}^{2P_{1/2}}=-\frac{m_{r}^{3}\alpha^{4}}{48m_X^{2}c^{2}}
\biggl(1+2\kappa_X+2\frac{m_X}{m_p}(1+\kappa_X)\biggr)
(1+A_{FS} r_p^2)+\ldots,
\end{equation}
where, 
$$A_{FS}=\frac{1}{4a_r^{2}}\left[-\left(1+\frac{\kappa_p}{(1-k^2)^{2}}
\right)-\left(1+\frac{\kappa_p}{1-k^2}\right)+\frac{\kappa_p}{k^{2}(1-k^2)^2}
\right], 
$$
with, $a_r = 1/(m_r \alpha)$ as before. The energy of the 
$2P_{3/2}$ level with form factor corrections is given by 
the relation $E_{FS}^{2P_{3/2}} = - (1/2) E_{FS}^{2P_{1/2}}$. 

\subsection{Hyperfine structure}
The hyperfine potential and the corresponding correction to the energy due to 
form factors in the case of muonic hydrogen has been discussed in detail in 
Ref. \cite{we3}. Here we give a brief description of this calculation 
for completeness. 
The $2S$ and $2P$ hyperfine energies relevant for this work are 
evaluated using the hyperfine potential, 
\begin{equation}\label{potrhfs}
\hat{V}_{hfs}(r) = {\alpha \mu_p\over 4 r^3 m_X m_p c^2}  
\biggl [ \mu_X \biggl \{ {3 (\mbox{\boldmath$\sigma$}_X 
\cdot \hat{\bf r} ) (\mbox{\boldmath$\sigma$}_p \cdot \hat{\bf r})} 
f_1(r)\, -\, {\mbox{\boldmath$\sigma$}_X \cdot \mbox{\boldmath$\sigma$}_p} 
f_2(r)\biggr \} 
+ 2 {{\bf L} \cdot \mbox{\boldmath$\sigma$}_p} f_3(r) \biggr ],  
\end{equation}
where, $\mu_X \, =\, 1\, +\, \kappa_X$,
$$f_1(r)\, =\, 1\, -\, e^{-mr} (1\,+\, mr)\, -\, {m^2 r^2 \over 6} 
\, e^{-mr} \, (3\, +\, mr),$$
$$f_2(r) = f_1(r) - (m^3 r^3/3)e^{-mr}\,\, {\rm and},$$
$$f_3(r) = 1\, -\, e^{-mr} (1\,+\, mr)\, 
-\, {m^2 r^2 \over 2} \, e^{-mr}.$$
The calculation of energies for states with $l = 0$ and $l \ne 0$ is done 
separately \cite{we3}. We will see in section IV that the hyperfine 
levels of relevance are 
$E_{hfs}^{2S_{1/2}^{f=1}}$, $E_{hfs}^{2S_{1/2}^{f=0}}$, 
$E_{hfs}^{2P_{3/2}^{f=2}}$ and $E_{hfs}^{2P_{3/2}^{f=1}}$. 
The complete expressions 
for any $n$ and $l$ can be found in \cite{we3}. 
For the $2S_{1/2}$ case, 
\begin{equation}\label{hyper2s}
E_{hfs}^{2S_{1/2}^{f=1}}=\frac{\alpha^4 m_r^3}{m_\mu
m_p}\frac{(1+\kappa_\mu)(1+\kappa_p)}{12}\left(1-\frac{6}{\sqrt{12}a_r}r_p+\frac{21}{12a_r^2}r_p^2-\frac{55}{12^{3/2}a_r^3}r_p^3+...\right),
\end{equation}
with $a_r = 1/(m_r \alpha)$. $E_{hfs}^{2S_{1/2}^{f=0}}=-3
E_{hfs}^{2S_{1/2}^{f=1}}$ and $(1/4) \Delta E_{hfs}^{2S} = 
E_{hfs}^{2S_{1/2}^{f=1}}$. 
Note the presence of the term linear in $r_p$. The sum of this correction 
and one arising from the Darwin term will contribute a small term 
linear in $r_p$ to the final expression for the transition energy, $\Delta$. 
Such a term 
linear in $r_p$ does not exist in the calculation of Pohl et al. \cite{pohl}. 
The energies of the $P$ levels are given as, 
\begin{eqnarray}
E_{hfs}^{2P_{3/2}^{f=1}}=
\left(\frac{C_1}{6}-\frac{5C_2}{6}-\frac{5C_3}{6}\right)
+\left(\frac{C_1}{3}+5C_2-\frac{5C_4}{6}\right)\frac{r_p^2}{12a^2}\\ \nonumber
+\left(-\frac{10C_1}{3}-\frac{50C_2}{3}-\frac{5C_5}{6}\right)\frac{r_p^3}{12^{3/2}a^3}+\ldots
,
\end{eqnarray}
where
\begin{eqnarray*}
 C_1&=&\frac{m_r^3\alpha^4}{m_\mu m_p}\frac{(1+\kappa_\mu)(1+\kappa_p)}{24},
\,\,\,
C_2=\frac{m_r^3\alpha^4}{m_\mu m_p}\frac{(1+\kappa_p)}{24},\\
C_3&=&\frac{m_r^{3}\alpha^{4}}{24m_\mu m_p}\left(\frac{m_\mu}{2m_p}\right)(1+2\kappa_p),\, \, \, 
C_4=\frac{m_r^{3}\alpha^{4}}{24m_\mu m_p}\left(\frac{m_\mu}{2m_p}\right)\left(-6(1+2\kappa_p)-3\frac{\kappa_p}{k^{2}}\right)\\
C_5&=&\frac{m_r^{3}\alpha^{4}}{24m_\mu
m_p}\left(\frac{m_\mu}{2m_p}\right)\left(20(1+2\kappa_p)-\frac{8\kappa_p}{(1-k^{2})^{2}}+\frac{8\kappa_p}{k^{3}(1-k^{2})^{2}}-\frac{12\kappa_p}{1-k^{2}}\right).
\end{eqnarray*}
\begin{eqnarray}
E_{hfs}^{2P_{3/2}^{f=2}}=\left(-\frac{C_1}{10}+\frac{C_2}{2}+\frac{C_3}{2}\right)+\left(3C_1-3C_2+\frac{C_4}{2}\right)\frac{r_p^2}{12a^2} \\ \nonumber
+\left(-14C_1+10C_2+\frac{C_5}{2}\right)\frac { r_p^3 } { 12^{3/2}a^3} +\ldots
\end{eqnarray}
with $C_1$, $C_2$, $C_3$ and $C_4$ the same as in the $2P_{3/2}^{f=1}$ case. 

The contribution of the $P$ levels to the finite size corrections are 
once again small.

\section{Radius of the proton}
In the next step, we evaluate the radius by calculating the energy 
corrections to expressions dependent on the proton radius $r_p$. 
Following a similar procedure as in \cite{pohl} conceptually, we perform 
a fit using the experimental value of the transition energy, 
$\Delta = E_{2P_{{3}/{2}}}^{f=2}-E_{2S_{{1}/{2}}}^{f=1}$ 
and the theoretical expressions for the corresponding energies (with finite 
size corrections) to determine the radius, $r_p$. 
In order to evaluate the energy difference, $\Delta = E_{2P_{{3}/{2}}}^{f=2}
-E_{2S_{{1}/{2}}}^{f=1}$, we need to find as in \cite{pohl}, 
\begin{eqnarray}\label{fshfsenergies}
E_{2S_{1/2}^{f=1}} &=& {1 \over 4 } \Delta E_{hfs}^{2S}, \\ \nonumber
E_{2P_{3/2}^{f=2}} &=& \Delta E_{LS} + \Delta E_{FS}^{2P_{3/2}} + {3 \over 8}
\Delta E_{hfs}^{2P_{3/2}},  \\ \nonumber
\Delta E_{LS} &=& E_{2P_{1/2}} - E_{2S_{1/2}}.
\end{eqnarray}
Here $f = j + s_p$ is the total angular momentum of the muon proton system 
with the muon total angular momentum,
$j = l + s_{\mu}$. 
The fine and hyperfine splittings respectively are, 
\begin{eqnarray}\label{finehyperfine}
\Delta E_{FS}^{2P_{3/2}} = E_{FS}^{2P_{3/2}} - E_{FS}^{2P_{1/2}}, \\ \nonumber 
\Delta E_{hfs}^{2P_{3/2}} = E_{hfs}^{2P_{3/2}^{f=2}} - 
E_{hfs}^{2P_{3/2}^{f=1}}.
\end{eqnarray} 
$\Delta E_{hfs}^{2S}$ is the hyperfine splitting of the $2S$ levels, i.e., 
$\Delta E_{hfs}^{2S} =E_{hfs}^{2S_{1/2}^{f=1}} - E_{hfs}^{2S_{1/2}^{f=0}} $. 
Such a fit in our case leads to the proton radius $r_p = 0.83112$ fm.
The correction to the 
Coulomb term evaluated from Eq. (\ref{coulombnew0}) gives 
$\Delta E_{Coul}^{2S_{1/2}} = 4.30248 r_p^2 - 0.020585 r_p^3 
\, {\rm meV}$, which forms 
the major contribution to the $r_p^2$ and $r_p^3$ terms to be discussed below.

\subsection{Coefficients of the $r_p^2$ and $r_p^3$ terms}
The value -5.2262 $r_p^2$ + 0.0347 $r_p^3$ meV
used in \cite{pohl} for  -$\Delta E_{Coul}^{2S_{1/2}}$
is based on the following equation taken from \cite{borie} 
which is based on Friar's formalism \cite{friar}: 
\begin{equation}\label{borie1}
\Delta E^{Borie} = {2 \alpha Z \over 3} \biggl ( {\alpha Z m_r \over n }
\biggr ) ^3 \, \, \biggl [\, <r^2>\,  - {\alpha Z m_r \over 2} <r^3>_{(2)} \, 
+ \, ... \biggr ] . 
\end{equation}
In this formalism, the
correction to the $1/r$ Coulomb potential takes the form
\begin{equation}\label{friar1}
\Delta V_c(r) = - Z \alpha \int \, d^3s \, \rho(s) \, 
\biggl (\, {1 \over |{\bf r} - {\bf s}| } - {1\over r} \, \biggr )  
\end{equation}
where $\rho(s)$ is the charge distribution in the proton. 
Using perturbation theory, the correction to the energy level is evaluated 
in \cite{friar} as (see Eq. (43a) in \cite{friar}),
\begin{equation}\label{friar2}
\Delta E \simeq {2 \pi Z \alpha \over 3} \, |\phi_n(0)|^2\, 
\biggl ( <r^2> - {Z\alpha\mu \over 2} <r^3>_{(2)} \, + \ldots\, \biggr ).
\end{equation}
In contrast to our formalism where we consider the full $r$ dependent
wave function and perform the integration, Ref. \cite{friar} 
considers the wave function
only at $r = 0$, as a result of which one is left with the integral 
$ < r^2 > \, = \, \int \, d^3r \, \rho(r) \, r^2$ in the energy correction. 
First we discuss the $< r^2 >$ dependent term and then go 
over to the term with $< r^3 >_{(2)}$. 

The energy corrections and hence the coefficients of the $r_p^2$ 
(writing $<r^2> = r_p^2$ as in \cite{pohl}) term as calculated from Friar's 
formalism will clearly differ from the present work due to the difference in 
Eqs (\ref{friar1}) and (\ref{coulpot}) for the two potentials. The 
finite size correction to the to Coulomb energy of the present work has been 
evaluated using the potential which is a Fourier transform of the 
first term in Eq.(\ref{potinqspace}), namely, 
$$U_{Coul}(\textbf{q})=4\pi e^2\,F_1^XF_1^p\Bigg
(-\frac{1}{\textbf{q}^2}\Biggr ).$$
If instead of taking just the Coulomb potential in momentum space, 
we decide to club one of the Darwin terms with it, we find
\begin{eqnarray}
U_{Coul}^{newdef}(\textbf{q})&=&4\pi e^2\biggl [ F_1^XF_1^p\biggl
(-\frac{1}{\textbf{q}^2} \biggr ) + F_1^X F_2^p\biggl
(\frac{1}{4m_p^2c^2}\biggr)\biggr]\\ \nonumber
&=& -4\pi \alpha\biggl [ {G_E^p({\bf q}^2)\over {\bf q}^2} \biggr ] . 
\end{eqnarray}
A Fourier transform of this potential with a dipole form factor 
leads to the following potential in coordinate space:
\begin{equation}
V_C^{newdef}=-\frac{\alpha}{r}\left[1-e^{-mr}
\left(1+\frac{mr}{2}\right)\right].
\end{equation}
The energy correction evaluated for the $2S$ state (i.e. n=2, $l$=0) using 
$\Delta V_C = V_C^{newdef} - (-\alpha/r)$ in 
$\Delta E_C = \int_0^{\infty}\Delta {V}_C |R_{nl}(r)|^2r^2dr$ is then
\begin{eqnarray}
\Delta E_{Coul}^{newdef}&=&\left(
\frac{1}{2a_r}\right)^{3}\frac{\alpha}{m^{2}}\Bigg[\left(\frac{ma_r}{1+ma_r}\right)^{2}\left(1+3F(-2,2;3;2/(1+ma_r))\right)+\\&&\newline\left(\frac{ma_r}{1+ma_r}\right)^{3}
\left(1+3F(-2,3;3;2/(1+ma_r))\right)\Bigg].
\end{eqnarray}
Expressing the hypergeometric function as a series expansion, truncating it at 
high orders in $\alpha$ and substituting $r_p^2 = 12/m^2$ leads to
\begin{equation}\label{newenergy}
\Delta E_{Coul}^{newdef} = 
\frac{\alpha^{4}m_r^{3}}{12}\left(r_p^{2}-\frac{5\alpha m_r}
{\sqrt{12}} r_p^3 + \ldots\right).
\end{equation}
The first term in Eq.(\ref{newenergy}) is exactly equal to that in 
(\ref{borie1}) for the $2S$ state. 
This means that in principle, taking one of the Darwin terms together 
with the Coulomb one in the Breit potential leads to an exact agreement with 
the coefficient of the $r_p^2$ term used in \cite{pohl}. However, it also 
means that there are cancellations due to the other Darwin terms which 
eventually leaves us with a smaller coefficient of the $r_p^2$ term  
than that used in \cite{pohl}. 

The second term in Eq.(\ref{borie1}) is the third moment of the 
convoluted proton charge density and is defined as,
\begin{equation}
<r^3>_{(2)}\, =\, \int\, d^3r\, r^3\, \rho_{(2)}(r)
\end{equation}
where the convoluted charge density is given by
\begin{equation}
\rho_{(2)} \, =\, \int\, d^3z \, \rho_{ch}(|{\bf z} - {\bf r}|)\, \rho_{ch}(z).
\end{equation}
Inserting the Fourier transform of the Sachs electric form factor $G_E(Q^2)$ 
for $\rho_{ch}$, one finds that
\begin{equation}
<r^3>_{(2)} = {48\over \pi}\, \int_0^{\infty}\, {dQ\over Q^4}
\, \biggl(\, G_E^2(Q^2)\, -\, 1\, +\, {Q^2\over3}<r^2>\, \biggr). 
\end{equation}
This third moment comes about due to the use of the smeared Coulomb 
potential (Eq. (\ref{friar1})) along with the perturbative expansion of the 
hydrogen wave functions. 
Using a dipole form factor $G_E$, one can show that 
$<r^3>_{(2)} = 35\sqrt{3} <r^2>^{3/2}$. This substitution in 
Eq. (\ref{borie1}) indeed leads to the factor 0.0347 $r_p^3$ in \cite{pohl}. 
Such a correction of order $\alpha^5$ which is proportional 
to the proton form factor squared is not included in the present work. 
Including a term with the proton form factor squared in the Breit potential
would correspond to a two photon exchange diagram which would obviously 
lead to a correction one order higher in $\alpha$ and hence would be 
quite small. 

\subsection{Darwin contribution to the Lamb 
shift and the fine and hyperfine corrections} 
The main contribution to the Lamb shift $\Delta E_{LS}$ comes from QED, namely, 
the Uehling potential. 
In the Nature paper \cite{pohl} using other methods 
\cite{zemach,friar} 
for including the proton structure corrections, the authors 
found 
\begin{equation}\label{natureLS}
\Delta E_{LS}^{Nature} = 206.0573(45) - 5.2262 r_p^2 + 0.0347 r_p^3\, 
{\rm meV},  
\end{equation}
where the first term includes the relativistic one loop vacuum polarization 
for a point nucleus plus other QED corrections and 
the second and third terms are due to finite size 
corrections (FSC). To evaluate $\Delta E_{LS}$ with FSC with 
the Breit potential approach, apart from 
$\Delta E_{Coul}^{2S_{1/2}} = 4.30248 r_p^2 - 0.020585 r_p^3 
\, {\rm meV}$, we need the corrections $\Delta E_{Coul}^{2P_{1/2}}$ 
and the $2S$ and $2P$ corrections to the Darwin terms too. The structure 
corrections to the Darwin and Coulomb terms of the $2P$ levels are 
negligibly small. 
If we use 
Eqs (\ref{darwin1}) and (\ref{darwin2}) for the $2S_{1/2}$ level, we get,
$E_{D}^{{\rm no}\, F_{1,2}(q^2)} = 13.768591 \, {\rm meV} $ and 
$E_{D}^{F_{1,2}(q^2=0)} = 14.4185121 \, {\rm meV}$.
Starting with Eq.(\ref{darwin3}) and following the same procedure of 
expanding the hypergeometric functions etc, we find for the $2S_{1/2}$ level, 
\begin{equation}\label{darwin4} 
E_D^{F_{1,2}(q^2)} =
14.418512 - 0.0793\:r_p + 0.0002613 \:r_p^2 - 
6.6 \times 10^{-6} \: r_p^3 + \ldots {\rm meV}.
\end{equation}
Note that the first term in the above equation is nothing but the
Darwin term with form factors taken at $q^2= 0$. 
Since the calculation of the Lamb shift as used in \cite{pohl} 
has been taken from
a relativistic calculation, in principle we do not need 
to take into account the Darwin term. However, that calculation was done 
without including proton form factors and hence 
we must include the ``correction" to 
the Darwin term due to form factors. Thus, we
subtract $E_{D}^{{\rm no}\, F_{1,2}(q^2)} = 13.768591 \, {\rm meV} $ 
from Eq. (\ref{darwin4}) and use 
\begin{equation}
\Delta E_{Darwin}^{2S_{1/2}} = 0.64992 - 0.0793 r_p + 0.0002613 r_p^2 - 
6.6 \times 10^{-7} \, r_p^3
\, {\rm meV}, 
\end{equation}
which is the correction to the Darwin term due to the proton form factors.  
Besides these, we also have, 
$\Delta E_{Coul}^{2P_{1/2}} = 3 \times 10^{-6} r_p^4 - 1.37 \times 
10^{-8} r_p^5 \, {\rm meV}$,
$\Delta E_{Darwin}^{2P_{1/2}} = 9.08 \times 10^{-8} r_p^2 - 4.4 \times 
10^{-10} r_p^3\, {\rm meV}$ which are however extremely small corrections. 

To evaluate $\Delta E_{LS}$, we start as in \cite{pohl} with the value 
206.0573 which is the sum of 24 terms from the table given in the 
supplementary material of \cite{pohl}. To this we add the FSC mentioned 
above for $2S_{1/2}$ and $2P_{1/2}$ levels and find 
$\Delta E_{LS} = E_{2P_{1/2}} - E_{2S_{1/2}}$ to be,
\begin{equation}\label{we3LS} 
\Delta E_{LS} = 205.40738 + 0.0793 r_p - 4.30274 r_p^2 + 0.020585 r_p^3\, 
{\rm meV}, 
\end{equation} 
which should be compared 
with $\Delta E_{LS}^{Nature}$ mentioned in (\ref{natureLS}) above. 
Our result for $\Delta E_{LS}$ with FSC includes an additional term 
in $r_p$ which is not present in $\Delta E_{LS}^{Nature}$. The coefficients 
of the $r_p^2$ and $r_p^3$ terms are however quite similar. 
The corrections to the $P$ levels are really tiny 
as expected. 

Finally, the proton size corrections to the hyperfine and fine structure 
energy levels are evaluated as explained in the previous section. 
The fine and hyperfine structure terms in (\ref{fshfsenergies}) and 
(\ref{finehyperfine})
are expressed in terms of the proton radius as 
follows:
\begin{eqnarray}\label{fshfswithff}
\Delta E_{FS}^{2P_{3/2}} &=& 8.34678 \,- \,4.26 \times 10^{-5} r_p^2\, + 
1.36 \times 10^{-7} \, r_p^3\, 
{\rm meV}\\ \nonumber
\Delta E_{hfs}^{2P_{3/2}} &=& 3.3912\, -\, 1.787 \times 10^{-5} r_p^2\, 
+ 5.45 \times 10^{-8} \, r_p^3\, 
{\rm meV}, 
\\ \nonumber
(1/4) \Delta E_{hfs}^{2S} &=&  5.708 - 0.0347 r_p + 0.0001 r_p^2 - 3.27 
\times 10^{-7}\, r_p^3\,  {\rm meV}. 
\end{eqnarray}
One can see that the FSC for fine structure are very small. The first term 
8.34678 meV is exactly the same as the sum of first two terms
given in Table I in the second reference of  
\cite{martynenko}. 

In the evaluation of the proton radius in
Pohl et al. \cite{pohl}, 
the values of the hyperfine splittings were taken from \cite{martynenko}, 
where the FSC for the $2S$ level were evaluated using the
Zemach method and those for the $2P$ case were not taken into account.
Their FSC (taken from Table II of the first reference in \cite{martynenko}) 
of order $\alpha^5$ and $\alpha^6$ sum to -$0.1535$ meV.
This correction is obtained using a Zemach radius of $R_Z = 1.022$ fm and
leads to
$\Delta E_{hfs}^{2S}$ = 22.8148 meV.
We note that using such a value of
$\Delta E_{hfs}^{2S}$ means that the information
about the radius of the proton (through the form factors or Zemach radius
$R_Z = 1.022$ fm) has already been included in the hyperfine energy which
is later used to extract the radius of the proton.
However, if we wish to remain within the spirit of the formalism used for 
$\Delta E_{LS}$, then we cannot take an input energy which already assumes
a certain radius of the proton. Hence, instead of using just a number 
for $\Delta E_{hfs}^{2S}$ (as done in \cite{pohl}) which already includes 
the FSC, we use the expressions given above in terms of $r_p$. 

Note that the above equations include 
the proton structure corrections {\it only}. 
Adding QED and other corrections to these levels 
as in Refs. [14,15] in \cite{pohl} and 
following a procedure similar to that 
in \cite{pohl} (thus differing only in the inclusion of the FSC), we obtain 
\begin{equation}\label{ourdelta} 
\Delta (= E_{2P_{3/2}}^{f=2} - E_{2S_{1/2}}^{f=1})  
= 209.16073 + 0.11388 r_p - 4.3029 r_p^2 + 0.020585 r_p^3 \, \, {\rm meV}, 
\end{equation}
as compared to, 
\begin{equation}\label{theirdelta} 
\Delta^{Nature} 
= 209.9779(49) - 5.2262 r_p^2 + 0.0347 r_p^3 \, \, {\rm meV},
\end{equation} 
obtained in \cite{pohl}.  

Eq. (\ref{ourdelta}) contains a term linear in $r_p$ which is not present 
in the expression (\ref{theirdelta}) of Pohl et al \cite{pohl}. 
This term arises partly 
due to the proton structure correction to the Darwin term and partly due to 
the $2S$ hyperfine energy level. 
For the $2S$ hyperfine level, this 
is a correction of order $\alpha^5$ as can be seen from 
Eq. (\ref{hyper2s}). In \cite{pohl}, such an order $\alpha^5$ correction 
of -0.1518 meV 
(see Table II in the first reference in 
\cite{martynenko} which was used in \cite{pohl})
has been included directly as a number while taking into account the energy of 
the $2S$ hyperfine level (see also the discussion in the paragraph below 
Eq.(\ref{fshfswithff})). The remaining part of the $r_p$ dependent term 
in our formalism arises from the form factor correction to the Darwin term
which contributes to the expression of the Lamb shift in (\ref{we3LS}). 
In \cite{pohl} the main contribution of 205.0282 meV 
to the Lamb shift arises from the relativistic one loop vacuum polarization
which has been evaluated in \cite{borie} for a point nucleus. 
Hence, though the relativistic effects 
(represented by the Darwin term in our formalism) are taken into account, the 
finite size corrections to this term are missing in \cite{pohl}.  

Taking the central value of the measured $\Delta = 206.2949(32)$ meV 
and replacing in the left hand side of Eq.(\ref{ourdelta}) 
leads to $r_p = 0.83112$ fm. If we compare 
this radius with $r_p = 0.84184(67)$ fm (obtained from (\ref{theirdelta})) 
in \cite{pohl} we see that an additional uncertainty arises due to the 
difference in the approaches used 
for including proton finite size effects. A detailed comparison of the 
present approach with those used in \cite{pohl} can be found in 
section IV of Ref. \cite{we3}. 

To conclude, we can say that it is gratifying to see that 
in spite of the differences 
in the approaches for calculating the FSC, we
obtain $r_p=0.83112$ fm which is not too far from the 
value of $r_p=0.84184(67)$ fm found in \cite{pohl}. 
It is however important to note that there exist different approaches 
\cite{we3,zemach,friar} for evaluating the finite size effects 
in literature which can lead to different results (and an additional 
uncertainty in the extracted radius). 
One should be cautious not to overestimate the accuracy 
in calculating the radius of the proton. 

\section*{Appendix: Evaluation of energies}
\appendix*
\setcounter{equation}{0}
The energies corresponding to the Coulomb, Darwin, fine structure and 
hyperfine structure terms in the Breit equation are evaluated by taking 
the expectation value of the corresponding potential using first order
time-independent perturbation theory. 
In general, for an operator $\hat{\bf A}$, the expectation value is, 
\begin{equation}
 \langle \hat{\textbf{A}}\rangle=\int r^2\:dr\:d\theta \:d\phi\:\Psi_{nlm_l}^*(r,\theta,\phi)\hat{\textbf{A}}\Psi_{nlm_l}(r,\theta,\phi)\label{B7}, 
\end{equation}
where, 
\begin{equation}
\Psi(r,\theta,\phi)=R_{nl}(r)Y_l^{m_l}(\theta, \phi), \label{B_4}
\end{equation}
with
\begin{equation}
R_{nl}(r)=\Bigg[\left(\frac{2}{na}\right)^3\frac{(n-l-1)!}{2n(n+l)!}\Bigg]^{1/2}e^{-r/na}\left(\frac{2r}{na}\right)^lL_{n-l-1}^{2l+1}(2r/na) \label{B5}
\end{equation}
being the radial functions. For operators which do not depend on angles, the 
expectation value reduces to calculating 
\begin{eqnarray}
 \langle \hat{\textbf{A}}\rangle&=&\left(\frac{2}{na}\right)^{2l+3}\frac{1}{2n2^{2(n-l-1)}}\sum_{j=0}^{n-l-1}{2(n-l-j-1)\choose n-l-j-1}\frac{(2j)!}{j!\Gamma(2l+j+2)}  \nonumber\\&&
 \times\int_0^\infty dr\textbf{A}e^{-2r/na}r^{2l+2}L_{2j}^{2(2l+1)}(4r/na).\label{B9}
\end{eqnarray}
The above integral can be evaluated analytically for the potentials considered 
in the present work. This amounts to evaluating the expectation values of 
terms of the type $e^{-mr}$, $r e^{-mr}$, $e^{-mr}/r$ etc. 
For example, 
 \begin{eqnarray}
  \left\langle e^{-mr} \right\rangle&=&\left(\frac{2}{na}\right)^{2l+3}\frac{1}{2n2^{2(n-l-1)}}\sum_{j=0}^{n-l-1}{2(n-l-j-1)\choose n-l-j-1}\frac{\Gamma(4l+2j+3)}{j!\Gamma(2l+j+2)}  \nonumber\\&&
  \frac{\Gamma(2l+3)}{\Gamma(4l+3)}\left(\frac{na}{2+mna}\right)^{2l+3}F\left(-2j,2l+3;4l+3;\frac{4}{2+mna}\right)\label{B12}
 \end{eqnarray}
 \begin{eqnarray}
  \left\langle \frac{e^{-mr}}{r} \right\rangle&=&\left(\frac{2}{na}\right)^{2l+3}\frac{1}{2n2^{2(n-l-1)}}\sum_{j=0}^{n-l-1}{2(n-l-j-1)\choose n-l-j-1}\frac{\Gamma(4l+2j+3)}{j!\Gamma(2l+j+2)}  \nonumber\\&&
  \frac{\Gamma(2l+2)}{\Gamma(4l+3)}\left(\frac{na}{2+mna}\right)^{2l+2}F\left(-2j,2l+2;4l+3;\frac{4}{2+mna}\right)\label{B13}
 \end{eqnarray}
 \begin{eqnarray}
  \left\langle \frac{e^{-mr}}{r^2} \right\rangle&=&\left(\frac{2}{na}\right)^{2l+3}\frac{1}{2n2^{2(n-l-1)}}\sum_{j=0}^{n-l-1}{2(n-l-j-1)\choose n-l-j-1}\frac{\Gamma(4l+2j+3)}{j!\Gamma(2l+j+2)}  \nonumber\\&&
  \frac{\Gamma(2l+1)}{\Gamma(4l+3)}\left(\frac{na}{2+mna}\right)^{2l+1}F\left(-2j,2l+1;4l+3;\frac{4}{2+mna}\right)\label{B14}
 \end{eqnarray}
 \begin{eqnarray}
  \left\langle \frac{e^{-mr}}{r^3} \right\rangle&=&\left(\frac{2}{na}\right)^{2l+3}\frac{1}{2n2^{2(n-l-1)}}\sum_{j=0}^{n-l-1}{2(n-l-j-1)\choose n-l-j-1}\frac{\Gamma(4l+2j+3)}{j!\Gamma(2l+j+2)}  \nonumber\\&&
  \frac{\Gamma(2l)}{\Gamma(4l+3)}\left(\frac{na}{2+mna}\right)^{2l}F\left(-2j,2l;4l+3;\frac{4}{2+mna}\right),\label{B15}
 \end{eqnarray}
with the expressions being dependent on the Gamma functions and hypergeometric 
functions. The latter can be expressed as, 
\begin{eqnarray}
 F(a,b;c;z)&=&1+\frac{ab}{1!c}z+\frac{a(a+1)b(b+1)}{2!c(c+1)}z^2+\ldots\nonumber\\
 &=&\sum_{n=0}^{\infty}\frac{(a)_n(b)_n}{(c)_n}\frac{z^n}{n!}.\label{B10}
\end{eqnarray}
The expectation value of the Coulomb potential with form factors 
in Eq.(\ref{coulpot}) for example leads to the Coulomb potential plus 
the finite size correction
to the Coulomb energy for any $n, l$ which is given by
\begin{eqnarray}
 &&\Delta E_{Coul}(n,l)=\alpha\left(\frac{2}{na_r}\right)^{2l+3}\frac{1}{2n2^{2(n-l-1)}\Gamma(4l+3)}\sum_{j=0}^{n-l-1}{2(n-l-j-1)\choose n-l-j-1}\frac{\Gamma(4l+2j+3)}{j!\Gamma(2l+j+2)}\nonumber\\&&
 \times\Bigg[\bigg(1+\frac{\kappa_p}{(1-k^2)^2}\bigg)\Gamma(2l+2)\bigg(\frac{na_r}{2+mna_r}\bigg)^{2l+2}F\bigg(-2j, 2l+2;4l+3;\frac{4}{2+mna_r}\bigg)\nonumber\\&&
 -\frac{\kappa_p}{(1-k^2)^2}\Gamma(2l+2)\bigg(\frac{na_r}{2+mkna_r}\bigg)^{2l+2}F\bigg(-2j, 2l+2;4l+3;\frac{4}{2+mkna_r}\bigg)\nonumber\\&&
 +\frac{m}{2}\bigg(1+\frac{\kappa_p}{1-k^2}\bigg)\Gamma(2l+3)\bigg(\frac{na_r}{2+mna_r}\bigg)^{2l+3}F\bigg(-2j, 2l+3;4l+3;\frac{4}{2+mna_r}\bigg)\Bigg].\label{4_3}
\end{eqnarray}

The energy corresponding to the Darwin term for example can be calculated 
by taking the expectation value of the potential in (\ref{darwinpot}) and 
leads to Eq.(\ref{darwin3}), namely, 
\begin{equation}\label{darwin33}
E_D^{F_{1,2}(q^2)}(n,l)=\frac{\alpha}{2m_X^2c^2}\frac{1}{n^3a_r^3} \Bigg[(1+2\kappa_X)G_{D1}(n,l) +\frac{m_X^2}{m_p^2}G_{D2}(n,l)\Bigg],
\end{equation}
where $G_{D1}(n,l)$ and $G_{D2}(n,l)$ are given as, 
\begin{eqnarray}
 &&G_{D1}(n,l)=\left(\frac{2}{na_r}\right)^{2l}\frac{1}{n2^{2(n-l-1)}}\sum_{j=0}^{n-l-1}{2(n-l-j-1)\choose n-l-j-1}\frac{\Gamma(4l+2j+3)}{j!\:\Gamma(2l+j+2)\Gamma(4l+3)}\nonumber\\&&
 \times\Bigg[\bigg(1+\frac{\kappa_p}{1-k²}\bigg)\frac{m^3}{2}\Gamma(2l+3)\Bigg(\frac{na_r}{2+mna_r}\Bigg)^{2l+3}F\left(-2j, 2l+3;4l+3;\frac{4}{2+mna_r}\right)\nonumber\\&&
 +\frac{m^2k^2\kappa_p}{(1-k^2)^2}\Gamma(2l+2)\left(\frac{na_r}{2+mna_r}\right)^{2l+2}F\left(-2j, 2l+2;4l+3;\frac{4}{2+mna_r}\right)\nonumber\\&&
 -\frac{m^2k^2\kappa_p}{(1-k^2)^2}\Gamma(2l+2)\left(\frac{na_r}{2+mkna_r}\right)^{2l+2}F\left(-2j, 2l+2;4l+3;\frac{4}{2+mkna_r}\right) \Bigg],\nonumber\\\nonumber\\&&
 G_{D2}(n,l)=\left(\frac{2}{na_r}\right)^{2l}\frac{1}{n2^{2(n-l-1)}}\sum_{j=0}^{n-l-1}{2(n-l-j-1)\choose n-l-j-1}\frac{\Gamma(4l+2j+3)}{j!\:\Gamma(2l+j+2)\Gamma(4l+3)}\nonumber\\&&
 \times\Bigg[\left(1+\kappa_p\left(\frac{1-2k^2}{1-k^2}\right)\right)\frac{m^3}{2}\Gamma(2l+3)\left(\frac{na_r}{2+mna_r}\right)^{2l+3}F\left(-2j, 2l+3;4l+3;\frac{4}{2+mna_r}\right)\nonumber\\&&
 -\frac{m^2k^2\kappa_p}{(1-k^2)^2}\Gamma(2l+2)\left(\frac{na_r}{2+mna_r}\right)^{2l+2}F\left(-2j, 2l+2;4l+3;\frac{4}{2+mna_r}\right)\nonumber\\&&
 +\frac{m^2k^2\kappa_p}{(1-k^2)^2} \Gamma(2l+2)\left(\frac{na_r}{2+mkna_r}\right)^{2l+2}F\left(-2j, 2l+2;4l+3;\frac{4}{2+mkna_r}\right)   \Bigg].\nonumber
\end{eqnarray}
Similarly, the fine structure potential with form factors gives rise to 
(\ref{fineenergy}) where, 
\begin{eqnarray}
 &&G_{FS}(n,l)=\left(\frac{2}{na_r}\right)^{2l+3}\frac{1}{2n2^{2(n-l-1)}}\sum_{j=0}^{n-l-1}{2(n-l-j-1)\choose n-l-j-1}\frac{\Gamma(4l+2j+3)}{j!\Gamma(2l+j+2)\Gamma(4l+3)}\nonumber\\&&
 \Bigg[-\Gamma(2l)\left(1+\frac{\kappa_p}{(1-k^2)^2}\right)\left(\frac{na_r}{2+mna_r}\right)^{2l}F\left(-2j,2l;4l+3;\frac{4}{2+mna_r}\right)\nonumber\\&&
 -\Gamma(2l+1)m\left(1+\frac{\kappa_p}{(1-k^2)^2}\right)\left(\frac{na_r}{2+mna_r}\right)^{2l+1}F\left(-2j,2l+1;4l+3;\frac{4}{2+mna_r}\right)\nonumber\\&&
 -\Gamma(2l+2)\frac{m^2}{2}\left(1+\frac{\kappa_p}{1-k^2}\right)\left(\frac{na_r}{2+mna_r}\right)^{2l+2}F\left(-2j,2l+2;4l+3;\frac{4}{2+mna_r}\right)\nonumber\\&&
  +\Gamma(2l)\frac{\kappa_p}{(1-k^2)^2}\left(\frac{na_r}{2+mkna_r}\right)^{2l}F\left(-2j,2l;4l+3;\frac{4}{2+mkna_r}\right)\nonumber\\&&
  +\Gamma(2l+1)mk\frac{\kappa_p}{(1-k^2)^2}\left(\frac{na_r}{2+mkna_r}\right)^{2l+1}F\left(-2j,2l+1;4l+3;\frac{4}{2+mkna_r}\right)\Bigg]. 
\end{eqnarray}

\end{document}